\documentclass[12pt]{iopart}

\usepackage{amssymb}
\usepackage{dcolumn}
\usepackage{bm}
\usepackage{color}
\usepackage{ulem}
\usepackage{graphicx}
\usepackage{braket}
\usepackage[perpage]{footmisc}

\begin{document}

\title[Square skyrmion crystal in centrosymmetric systems]{Square skyrmion crystal in centrosymmetric systems with locally inversion-asymmetric layers}

\author{Satoru Hayami}

\address{Department of Applied Physics, University of Tokyo, Bunkyo, Tokyo 113-8656, Japan \\
Faculty of Science, Hokkaido University, Sapporo 060-0810, Japan
}
\ead{hayami@phys.sci.hokudai.ac.jp}
\vspace{10pt}

\begin{abstract}
We investigate an instability toward a square-lattice formation of magnetic skyrmions in centrosymmetric layered systems. 
By focusing on a bilayer square-lattice structure with the inversion center at the interlayer bond instead of the atomic site, we numerically examine the stability of the square skyrmion crystal based on an effective spin model with the momentum-resolved interaction in the ground state through the simulated annealing. 
As a result, we find that a layer-dependent staggered Dzyaloshinskii-Moriya interaction built in the lattice structure becomes the origin of the square skyrmion crystal in an external magnetic field irrespective of the sign of the interlayer exchange interaction. 
The obtained square skyrmion crystal is constituted of the skyrmion crystals with different helicities in each layer due to the staggered Dzyaloshinskii-Moriya interaction. 
Furthermore, we show that the interplay between the staggered Dzyaloshinskii-Moriya interaction and the interlayer exchange interaction gives rise to a double-$Q$ state with a uniform component of the scalar chirality in the low-field region. 
The present results provide another way of stabilizing the square skyrmion crystal in centrosymmetric magnets, which will be useful to explore further exotic topological spin textures. 
\end{abstract}

%
%
%
%
%


\section{Introduction}
\label{sec:intro}

One of the central issues in the field of magnetism is to discover materials hosting nontrivial topological spin textures since they give rise to unconventional physical phenomena arising from giant emergent electromagnetic fields, such as the topological Hall effect~\cite{Bruno_PhysRevLett.93.096806, Neubauer_PhysRevLett.102.186602,  Kanazawa_PhysRevLett.106.156603,Hayami_PhysRevB.94.024424,nakazawa2018topological} and the nonlinear optics~\cite{nagaosa2019emergent,yokouchi2020emergent, Ieda_PhysRevB.103.L100402, kitaori2021emergent, kurebayashi2021electromagnetic}. 
A typical example to possess such topological spin textures is a magnetic skyrmion with a two-dimensional topological spin texture~\cite{Bogdanov89, Bogdanov94,rossler2006spontaneous,nagaosa2013topological}, which has been studied in both theory and experiment since the first observation of its crystal form referred to as the magnetic skyrmion crystal (SkX) in the cubic chiral magnet MnSi in 2009~\cite{Muhlbauer_2009skyrmion}. 
Subsequently, the emergence of the SkXs has been clarified in various noncentrosymmetric magnets~\cite{yu2010real,yu2011near,heinze2011spontaneous,seki2012observation,Adams2012,Seki_PhysRevB.85.220406,tokunaga2015new,kezsmarki_neel-type_2015,karube2016robust,Li_PhysRevB.93.060409,Kurumaji_PhysRevLett.119.237201,nayak2017discovery,peng2020controlled,Tokura_doi:10.1021/acs.chemrev.0c00297}, where it was shown that the competition between the ferromagnetic (FM) exchange interaction and the Dzyaloshinskii-Moriya (DM) interaction, the latter of which originates from relativistic spin-orbit coupling in inversion-asymmetric lattice structures~\cite{dzyaloshinsky1958thermodynamic,moriya1960anisotropic}, in an external magnetic field is a key essence to stabilize the SkX from the theoretical aspect~\cite{rossler2006spontaneous,Yi_PhysRevB.80.054416, Binz_PhysRevLett.96.207202, Binz_PhysRevB.74.214408}. 
Thus, the absence of the inversion symmetry in the lattice structures is an important ingredient to realize the topological spin textures, which provides a guideline for their exploration. 
Indeed, the extensive searches based on this idea revealed the appearance of further unconventional short-period SkX in EuPtSi~\cite{kakihana2018giant,kaneko2019unique,kakihana2019unique,tabata2019magnetic} and a hedgehog lattice characterized by a three-dimensional topological spin texture in MnSi$_{1-x}$Ge$_{x}$~\cite{Binz_PhysRevB.74.214408,Park_PhysRevB.83.184406,Yang2016,tanigaki2015real,kanazawa2017noncentrosymmetric,fujishiro2019topological}, which are accounted for by introducing multiple-spin interactions in addition to the DM interaction~\cite{heinze2011spontaneous, Hayami_PhysRevLett.121.137202,brinker2019chiral, Okumura_PhysRevB.101.144416, Mankovsky_PhysRevB.101.174401,paul2020role,Brinker_PhysRevResearch.2.033240,lounis2020multiple,grytsiuk2020topological, Kathyat_PhysRevB.103.035111,hayami2021field,Mendive-Tapia_PhysRevB.103.024410,hayami2021topological, Hayami_PhysRevB.104.094425, Kato_PhysRevB.104.224405}. 

Meanwhile, a way of stabilizing the SkX and hedgehog lattice has been theoretically established even in centrosymmetric magnets. 
There have been so far several mechanisms for such topological spin textures, such as the frustrated exchange interaction~\cite{Okubo_PhysRevLett.108.017206,leonov2015multiply,Lin_PhysRevB.93.064430,Hayami_PhysRevB.93.184413,Hayami_PhysRevB.94.174420,batista2016frustration,Lin_PhysRevLett.120.077202,Hayami_PhysRevB.103.224418,hayami2022skyrmion,Aoyama_PhysRevB.105.L100407}, the Ruderman-Kittel-Kasuya-Yosida interaction~\cite{Wang_PhysRevLett.124.207201,Mitsumoto_PhysRevB.104.184432,Mitsumoto_PhysRevB.105.094427}, the multiple-spin interaction~\cite{Akagi_PhysRevLett.108.096401,Hayami_PhysRevB.90.060402,Ozawa_PhysRevLett.118.147205,Hayami_PhysRevB.95.224424,Hayami_PhysRevB.99.094420,Simon_PhysRevMaterials.4.084408,hayami2020multiple,hayami2021topological,Eto_PhysRevB.104.104425,Hayami_10.1088/1367-2630/ac3683,hayami2021phase,eto2022low}, and the crystal-dependent two-spin anisotropic interaction~\cite{Michael_PhysRevB.91.155135,Rousochatzakis2016,amoroso2020spontaneous,yambe2021skyrmion, Hayami_PhysRevB.103.024439, Hayami_PhysRevB.103.054422, Utesov_PhysRevB.103.064414, Wang_PhysRevB.103.104408, amoroso2021tuning,yambe2022effective}. 
These studies provide a deep understanding of the microscopic origins of topological spin textures in centrosymmetric magnets, which have recently been observed in Gd$_2$PdSi$_3$~\cite{Saha_PhysRevB.60.12162,kurumaji2019skyrmion,sampathkumaran2019report,Hirschberger_PhysRevB.101.220401,Kumar_PhysRevB.101.144440,Spachmann_PhysRevB.103.184424,paddison2022magnetic,bouaziz2022fermi} and  Gd$_3$Ru$_4$Al$_{12}$~\cite{hirschberger2019skyrmion,Hirschberger_10.1088/1367-2630/abdef9} hosting the triangular-lattice SkX, GdRu$_2$Si$_2$~\cite{khanh2020nanometric,Yasui2020imaging,khanh2022zoology} and EuAl$_4$~\cite{Shang_PhysRevB.103.L020405,kaneko2021charge,Zhu_PhysRevB.105.014423,takagi2022square} hosting the square-lattice SkX, and SrFeO$_3$ hosting the hedgehog lattice~\cite{Ishiwata_PhysRevB.84.054427,Ishiwata_PhysRevB.101.134406,Rogge_PhysRevMaterials.3.084404,Onose_PhysRevMaterials.4.114420}. 

Under these circumstances, an interesting situation to realize the triangular SkX in centrosymmetric lattice structures has been investigated by focusing on the role of the sublattice-dependent DM interaction~\cite{Diaz_hysRevLett.122.187203, Fang_PhysRevMaterials.5.054401, Hayami_PhysRevB.105.014408,lin2021skyrmion,Hayami_PhysRevB.105.184426}. 
Such a situation occurs when considering the multi-sublattice systems in centrosymmetric lattice structures, such as the zigzag~\cite{Yanase_JPSJ.83.014703,Hayami_doi:10.7566/JPSJ.84.064717,Hayami_doi:10.7566/JPSJ.85.053705,Sumita_PhysRevB.93.224507,cysne2021orbital, Suzuki_PhysRevB.105.075201,Yatsushiro_PhysRevB.105.155157}, honeycomb~\cite{Kane_PhysRevLett.95.226801,Hayami_PhysRevB.90.081115,yanagi2017optical,Yanagi_PhysRevB.97.020404,Hayami_PhysRevB.105.014404}, diamond~\cite{Fu_PhysRevLett.98.106803,Hayami_PhysRevB.97.024414,Ishitobi_doi:10.7566/JPSJ.88.063708}, and bilayer~\cite{hitomi2014electric,hitomi2016electric,yatsushiro2020odd,Yatsushiro_PhysRevB.102.195147} structures, where the inversion center lies at the bond center between the different sublattices while there is no local inversion symmetry at atomic sites~\cite{zhang2014hidden,Hayami_PhysRevB.90.024432,Fu_PhysRevLett.115.026401,Razzoli_PhysRevLett.118.086402,hayami2016emergent,gotlieb2018revealing,Huang_PhysRevB.102.085205,Ishizuka_PhysRevB.98.224510}. 
Since the sublattice-dependent DM interaction ubiquitously appears in the centrosymmetric systems to possess the magnetic ions located at the Wyckoff positions without the inversion center, its mechanism extends the scope of the candidate materials hosting the topological spin textures. 

In the present study, we further investigate the possibility of the SkX in the centrosymmetric lattice systems consisting of locally inversion-asymmetric layers. 
In particular, we aim at elucidating a new mechanism of the square SkX by considering a bilayer square-lattice system, as its stabilization mechanism is limited compared to the triangular SkX owing to the different nature of the multiple-$Q$ superposition: The former is characterized by the double-$Q$ spiral superposition with two orthogonal ordering vectors $\bm{Q}_1$ and $\bm{Q}_2$, i.e., $\bm{Q}_1 + \bm{Q}_2 \neq \bm{0}$, while the latter is represented by the triple-$Q$ one with $\bm{Q}_1$, $\bm{Q}_2$, and $\bm{Q}_3$ satisfying $\bm{Q}_1+ \bm{Q}_2 + \bm{Q}_3=\bm{0}$. 
An effective coupling by the multiple-$Q$ modulation in the form of $(\bm{S}_{\bm{0}}\cdot \bm{S}_{\bm{Q}_1})(\bm{S}_{\bm{Q}_2}\cdot \bm{S}_{\bm{Q}_3})$ appears in the free energy only in the latter triple-$Q$ case, which results in the defferent stabilization tendency between the sqaure and triangular SkXs. 
In fact, the stabilization of the centrosymmetric square SkX has been achieved by considering the further effect of the bond-dependent anisotropic exchange interaction or the dipolar interaction in addition to the frustrated exchange interaction in insulating magnets~\cite{Utesov_PhysRevB.103.064414, Wang_PhysRevB.103.104408} and the multiple-spin interaction in magnetic metals~\cite{Christensen_PhysRevX.8.041022,Hayami_doi:10.7566/JPSJ.89.103702,Hayami_PhysRevB.103.024439,Hayami_PhysRevB.105.104428,Steffensen_PhysRevResearch.4.013225}, which reproduces the multiple-$Q$ instabilities observed in GdRu$_2$Si$_2$~\cite{khanh2020nanometric,Yasui2020imaging,khanh2022zoology}. 
Besides, the scenario based on the competing interactions in momentum space has recently been proposed, which might explain the multiple SkX phases including the square SkX in EuAl$_4$~\cite{takagi2022square,hayami2022multiple,hayami2022rectangular,Hayami_PhysRevB.105.174437}. 

By analyzing an effective spin model incorporating the effect of the staggered DM interaction and the interlayer exchange interaction and performing the simulated annealing, we find that the square SkX consisting of the layer-dependent SkXs with different helicities is stabilized on the bilayer square lattice. 
Although the square SkX with a quantized integer skyrmion number appears for both FM and antiferromagnetic (AFM) interlayer exchange interactions, the resultant spin textures are different from each other. 
The skyrmion core positions on the different layers are different (the same) for the FM (AFM) interlayer exchange interactions. 
In addition to the square SkX, we find that the single-$Q$ spiral spin state in the low-field region is modulated so as to have a uniform scalar chirality in the presence of the interlayer exchange interaction. 
We also discuss similarities and differences of the results in the bilayer triangular-lattice system~\cite{Hayami_PhysRevB.105.014408}. 
Our result provides a new lattice structure hosting the square SkX in centrosymmetric magnets. 

The rest of the paper is organized as follows. 
In Sec.~\ref{sec: set up}, we introduce the effective spin model including the staggered DM interaction on the layered structure. 
We also present the numerical method based on the simulated annealing. 
In Sec.~\ref{sec: results}, we discuss the main results in the present paper. 
After showing the result in the single-layer system, we discuss the instability toward the square SkX while changing the sign and the magnitude of the interlayer exchange interaction. 
We present the details of the spin and chirality textures in real and momentum spaces obtained by the simulated annealing. 
Section~\ref{sec: summary} is devoted to a summary.

\section{Set up}
\label{sec: set up}

\subsection{Spin model}
\label{sec: Spin model}

\begin{figure}[htb!]
\begin{center}
\includegraphics[width=0.5 \hsize]{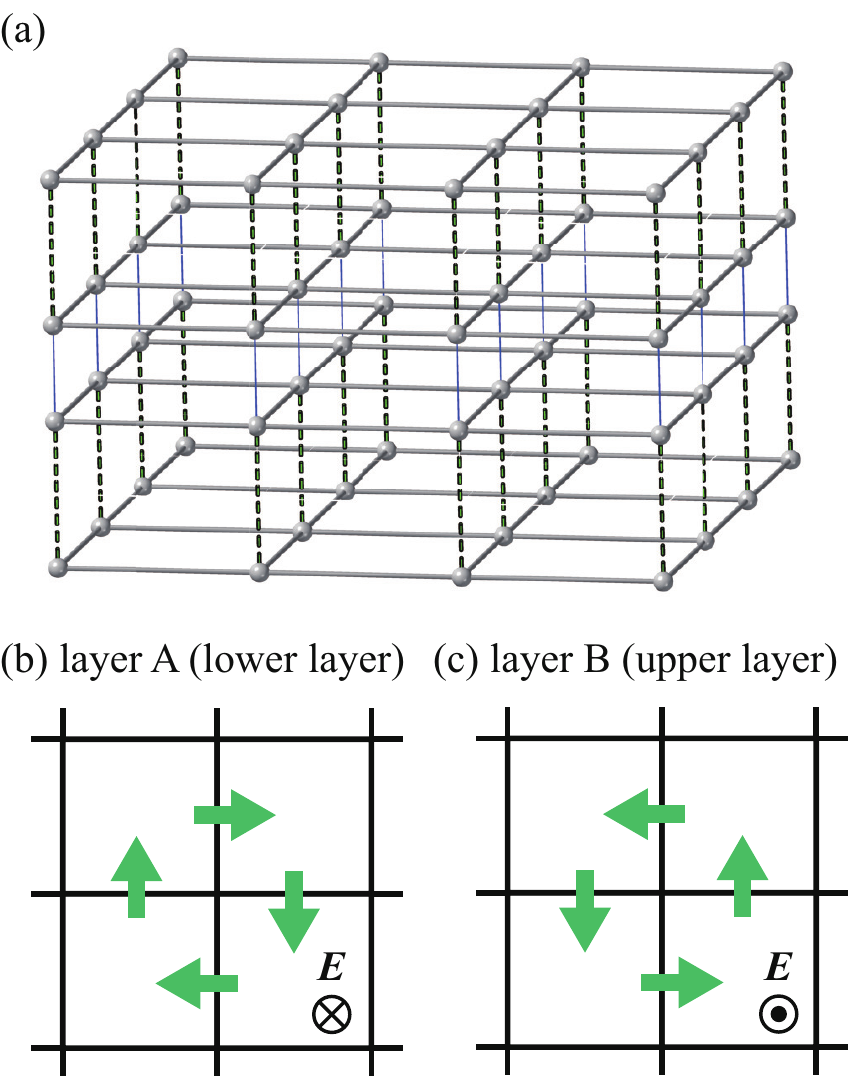} 
\caption{
\label{Fig: lattice}
(a) The layered square-lattice system with the staggered DM interaction, where the magnetic instabilities for the middle bilayer are investigated. 
(b), (c) The bilayer square-lattice structure consisting of (b) layer A and (c) layer B, which are stacked along the $z$ direction. 
The magnetic sites in each layer are affected by the local crystalline electric field $\bm{E}$ along the opposite directions, where the green arrows represent the DM vectors. 
}
\end{center}
\end{figure}

The staggered DM interaction appears in the lattice structure where there is no inversion center at the magnetic ions but there is an inversion center at the bond. 
We show an example satisfying such a condition in figure~\ref{Fig: lattice}(a); the two different square-lattice planes lie in the $xy$ plane and they are stacked along the $z$ direction with alternative bonds denoted by the dashed green lines and the thin blue lines. 
This lattice structure corresponds to the 2$g$ site under the space group \#123, where the site symmetry is $C_{4v}$; the inversion center is located at the bond center along the $z$ direction. 

We investigate the instability toward the multiple-$Q$ states in such a system with the staggered DM interaction.  
For simplicity, we here consider a single bilayer square-lattice system by extracting the middle bilayer connected by the thin blue lines from the other layers in figure~\ref{Fig: lattice}(a). 
We label the lower and upper layers in the targeting middle bilayer as layer A and layer B, respectively. 
As there is no inversion center in the intralayer, a crystalline electric field occurs locally on each layer along the $z$ direction. 
Owing to the inversion symmetry at the bond center along the $z$ direction, the A and B sites feel the electric field with the same magnitude but in the opposite direction. 
Accordingly, the staggered DM vector is induced in the directions perpendicular to the intralayer bond direction and the $z$ direction in the opposite way for layers A and B, as shown in figures~\ref{Fig: lattice}(b) and \ref{Fig: lattice}(c).
 
Then, the spin model on the bilayer square lattice is given by 
 \begin{eqnarray}
\label{eq: Ham}
\mathcal{H}=&\sum_{\eta}\mathcal{H}^{\perp}_{\eta}+\mathcal{H}^{\parallel}+\mathcal{H}^{{\rm Z}}, \\
\label{eq: Ham_perp}
\mathcal{H}^{\perp}_{\eta}=&  -\sum_{ij} \left[ J_{ij} \bm{S}_{i} \cdot \bm{S}_{j}+    \bm{D}_{ij}^{(\eta)} \cdot  (\bm{S}_{i} \times \bm{S}_{j}) \right], \\
\label{eq: Ham_parallel}
\mathcal{H}^{\parallel}=& J_{\parallel} \sum_{i} \bm{S}_i \cdot \bm{S}_{i+\hat{z}},\\
\label{eq: Ham_Zeeman}
\mathcal{H}^{{\rm Z}}=&-H \sum_i S_i^z, 
\end{eqnarray}
where $\mathcal{H}$ represents the total Hamiltonian in the system consisting of three terms; $\bm{S}_i$ is the classical localized spin at site $i$ with $|\bm{S}_i|=1$.  
The first term $\mathcal{H}^{\perp}_{\eta}$ in (\ref{eq: Ham}) represents the intralayer Hamiltonian for layer $\eta={\rm A}, {\rm B}$, which includes the layer-independent exchange interaction $J_{ij}$ and the layer-dependent staggered DM interaction $\bm{D}_{ij}^{(\eta)}$ ($\bm{D}_{ij}^{({\rm A})}=-\bm{D}_{ij}^{({\rm B})}$ and $|\bm{D}_{ij}^{({\rm A})}|=|\bm{D}_{ij}^{({\rm B})}| \equiv D_{ij}$). 
It is noted that the DM vector has only the $xy$ component from the symmetry, as shown in figures~\ref{Fig: lattice}(b) and \ref{Fig: lattice}(c). 
The second term $\mathcal{H}^{\parallel}$ in (\ref{eq: Ham}) represents the interlayer Hamiltonian, where the positive and negative signs of $J_{\parallel}$ correspond to the AFM and FM interlayer exchange interactions, respectively. 
The last term $\mathcal{H}^{{\rm Z}}$ in (\ref{eq: Ham}) represents the Zeeman coupling in an external magnetic field along the $z$ direction. 
We neglect long-range dipolar interactions for simplicity.

For the model in (\ref{eq: Ham}), we investigate the stability of the square SkX characterized by a superposition of double-$Q$ spiral states. 
For that purpose, we suppose a finite-$Q$ magnetic instability as a result of the competition between the intralayer exchange interaction $J_{ij}$ and the DM interaction $\bm{D}^{(\eta)}_{ij}$ in each layer. 
In such an assumption, the intralayer Hamiltonian $\mathcal{H}^{\perp}_{\eta}$ is simplified as 
\begin{eqnarray}
\label{eq: Ham_perp2}
\tilde{\mathcal{H}}^{\perp}_\eta=&  -\sum_{\nu}  \Big[ J_{\bm{Q}_\nu} \bm{S}^{(\eta)}_{\bm{Q}_{\nu}} \cdot \bm{S}^{(\eta)}_{-\bm{Q}_{\nu}}+ i   \bm{D}^{(\eta)}_{\bm{Q}_\nu} \cdot ( \bm{S}^{(\eta)}_{\bm{Q}_{\nu}} \times \bm{S}^{(\eta)}_{-\bm{Q}_{\nu}}) \Big],  
\end{eqnarray}
where $\bm{S}^{(\eta)}_{\bm{Q}_{\nu}}$ is obtained from the Fourier transform of $\bm{S}_i$ at wave vector $\bm{Q}_\nu$ for layer $\eta$; $J_{\bm{Q}_\nu}$ and $\bm{D}^{(\eta)}_{\bm{Q}_\nu}$ are also the Fourier transforms of $J_{ij}$ and $\bm{D}_{ij}^{(\eta)}$, respectively. 
The summation of $\nu$ is taken over the dominant $\bm{Q}_\nu$ among the wave vector $\bm{q}$ in the Brillouin zone, which is determined by the minimum eigenvalues of the Fourier transform of $\mathcal{H}^{\perp}_{\eta}$ in (\ref{eq: Ham_perp}): $-\sum_{\bm{q}} [J^{(\eta)}_{\bm{q}} \bm{S}^{(\eta)}_{\bm{q}} \cdot \bm{S}^{(\eta)}_{-\bm{q}}+i   \bm{D}^{(\eta)}_{\bm{q}} \cdot ( \bm{S}^{(\eta)}_{\bm{q}} \times \bm{S}^{(\eta)}_{-\bm{q}})]$. 
Although the dominant $\bm{Q}_\nu$ components and the magnitudes of $J_{\bm{Q}_\nu}$ and $\bm{D}^{(\eta)}_{\bm{Q}_\nu}$ are determined by the real-space interactions in (\ref{eq: Ham_perp}), we take them as phenomenological parameters for simplicity. 
From the fourfold rotational symmetry of the square-lattice system, there are four minima in $\bm{q}$ space; we take $\bm{Q}_1=(\pi/3,0)$, $\bm{Q}_2=(0,\pi/3)$, $\bm{Q}_3=(-\pi/3,0)$, and $\bm{Q}_4=(0,-\pi/3)$, where we suppose that $\bm{Q}_{\nu}$ lies on the high-symmetry lines on the Brillouin zone. 
For a given set of $\bm{Q}_\nu$, $J_{\bm{Q}_\nu} \equiv J$ and $D^y_{\bm{Q}_1} = -D^x_{\bm{Q}_2}=-D^y_{\bm{Q}_3}=D^x_{\bm{Q}_4} \equiv D$ owing to the fourfold rotational symmetry. 
By considering the multiple-$Q$ magnetic instability at low temperatures, we neglect the contributions from the other $\bm{q}$ components in the interactions. 
In other words, we do not consider the possibility of the triangular SkX, which might appear in the square-lattice system~\cite{Lin_PhysRevB.91.224407}.
In the following, we set $J=1$ and $D=0.2$, which stabilizes the square SkX in the limit of $J_{\parallel}=0$, as will be shown in Sec.~\ref{sec: single layer}. 
For fixed $J$ and $D$, we examine its stability while changing $J_{\parallel}$ and $H$. 

\subsection{Simulated annealing}
\label{sec: Simulated annealing}

The low-temperature spin configuration of the model $\tilde{\mathcal{H}}=\sum_{\eta}\tilde{\mathcal{H}}^{\perp}_\eta+\mathcal{H}^{\parallel}+\mathcal{H}^{{\rm Z}}$ on the bilayer square lattice is calculated by performing the simulated annealing based on the standard Metropolis local updates in real space. 
The total number of spins is taken as $N=2\times L^2$ with $L=48$. 
Starting from the high-temperature ranged from 1 to 10, i.e., $1\leq T_0 \leq 10$ ($T_0$ is the initial temperature), we gradually reduce the temperature with a ratio $T_{n+1}=\alpha T_n$ in each Monte Carlo sweep, where $T_n$ is the $n$th-step temperature and $\alpha=0.999999$. 
The decrease of the temperature is continued until it reaches the final temperature $T=0.001$. 
After reaching the final temperature, we perform $10^5$-$10^6$ Monte Carlo sweeps for thermalization and measurements. 
The simulations are independently performed for a given parameter set of $J_{\parallel}$ and $H$. 

To identify the magnetic phases from the spin configurations obtained by the simulated annealing, we calculate the spin and chirality structure factors. 
The spin structure factor for layer $\eta=\mathrm{A}, \mathrm{B}$ and spin-component $\alpha=x,y,z$ is given by 
\begin{equation}
S_{\eta}^\alpha(\bm{q})= \frac{1}{L^2} \sum_{i,j \in \eta} S^{\alpha}_i S^{\alpha}_j e^{i\bm{q}\cdot (\bm{r}_i-\bm{r}_j)}, 
\end{equation}
where the site indices $i$ and $j$ are taken for the same layer. 
We also calculate $S_{\eta}^{xy}(\bm{q})=S_{\eta}^x(\bm{q})+S_{\eta}^y(\bm{q})$.
The net magnetization for each layer is given by $M^\alpha_{\eta}=(1/L^2)\sum_{i \in \eta}S^{\alpha}_{i}$; we also define the inplane component of the uniform magnetization $M^{xy}_{\eta}=\sqrt{(M^x_\eta)^2+(M^y_\eta)^2}$.  
Meanwhile, the spin scalar chirality for layer $\eta$ is represented by 
\begin{eqnarray}
\chi^{\rm sc}_{\eta} &=& \frac{1}{L^2} \sum_{i \in \eta} \sum_{\delta=\pm1}\chi_i, \\
\chi_i&=&\bm{S}_i \cdot (\bm{S}_{i+\delta \hat{x}}\times \bm{S}_{i+\delta \hat{y}}),
\end{eqnarray}
where $\hat{x}$ ($\hat{y}$) is the unit vector in the $x$ ($y$) direction on the square lattice~\cite{Yi_PhysRevB.80.054416}. 
The total scalar chirality is given by $\chi^{\rm sc}=\chi^{\rm sc}_{\rm A}+ \chi^{\rm sc}_{\rm B}$, which is a measure of the topological Hall effect. 
The skyrmion number is calculated for each layer by using the scalar chirality as~\cite{BERG1981412}
\begin{equation}
\label{eq:nsk_num}
n^{(\eta)}_{\rm sk}=\frac{1}{2\pi N_m}\sum_{i \in \eta} \sum_{\delta=\pm 1}  \tan^{-1} \frac{\bm{S}_i \cdot (\bm{S}_j \times \bm{S}_k)}{1+\bm{S}_i \cdot \bm{S}_j+\bm{S}_j \cdot \bm{S}_k+\bm{S}_k \cdot \bm{S}_i},
\end{equation} 
where $N_m$ is the number of magnetic unit cell in the system, and $j=i+\delta \hat{x}$ and $k=i+\delta \hat{y}$; the range of the arctangent is set as $[-\pi, \pi)$. 
For example, $n^{(\eta)}_{\rm sk}=-1$ when the SkX appears for layer $\eta$.
The scalar chirality structure factor is calculated by using $\chi_i$ as 
\begin{equation}
\label{eq:chiralstructurefactor}
S^{\chi}_{\eta}(\bm{q})= \frac{1}{L^2}\sum_{\delta=\pm 1}\sum_{i,j \in \eta}  \chi_{i}
\chi_{j} e^{i \bm{q}\cdot (\bm{r}_i-\bm{r}_j)}.  
\end{equation}

\section{Results}
\label{sec: results}
In this section, we show the stability of the square SkX in the model $\tilde{\mathcal{H}}=\sum_\eta\tilde{\mathcal{H}}^{\perp}_\eta+\mathcal{H}^{\parallel}+\mathcal{H}^{{\rm Z}}$. 
First, we discuss the result in the single-layer case against the magnetic field in Sec.~\ref{sec: single layer}. 
Then, we discuss the result in the bilayer case while changing the magnetic field and interlayer exchange interaction in Sec.~\ref{sec: bilayer}. 

\subsection{Single-layer case}
\label{sec: single layer}

\begin{figure}[htb!]
\begin{center}
\includegraphics[width=0.4 \hsize]{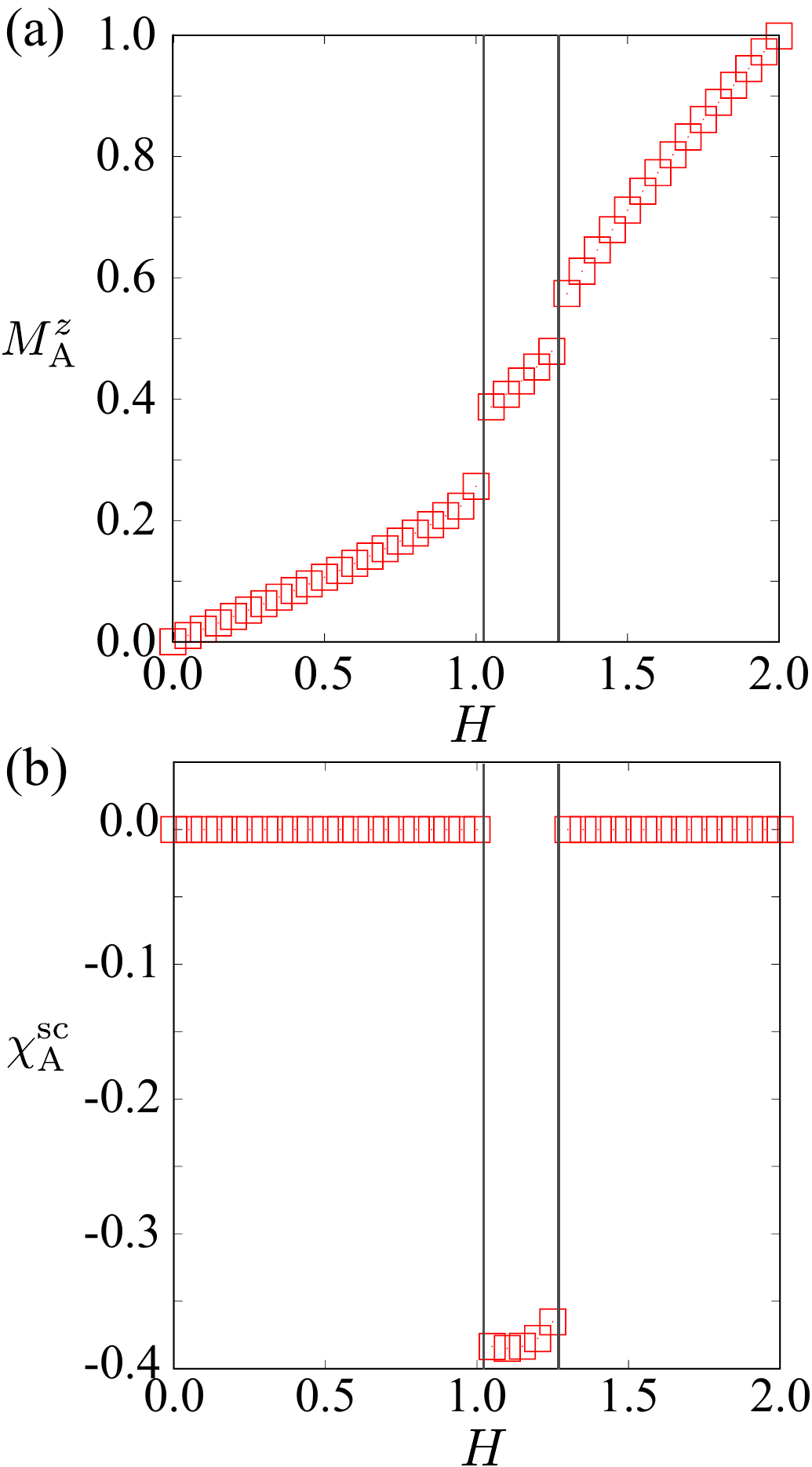} 
\caption{
\label{Fig: Mag_chirality_SL}
$H$ dependences of (a) the magnetization $M^z_\eta$ and (b) the scalar chirality $\chi^{\rm sc}_\eta$ for signle layer A at $J_{\parallel}=0$. 
The solid vertical lines represent the phase boundaries between the SkX and the other magnetic phases. 
}
\end{center}
\end{figure}

\begin{figure*}[htb!]
\begin{center}
\includegraphics[width=0.6 \hsize]{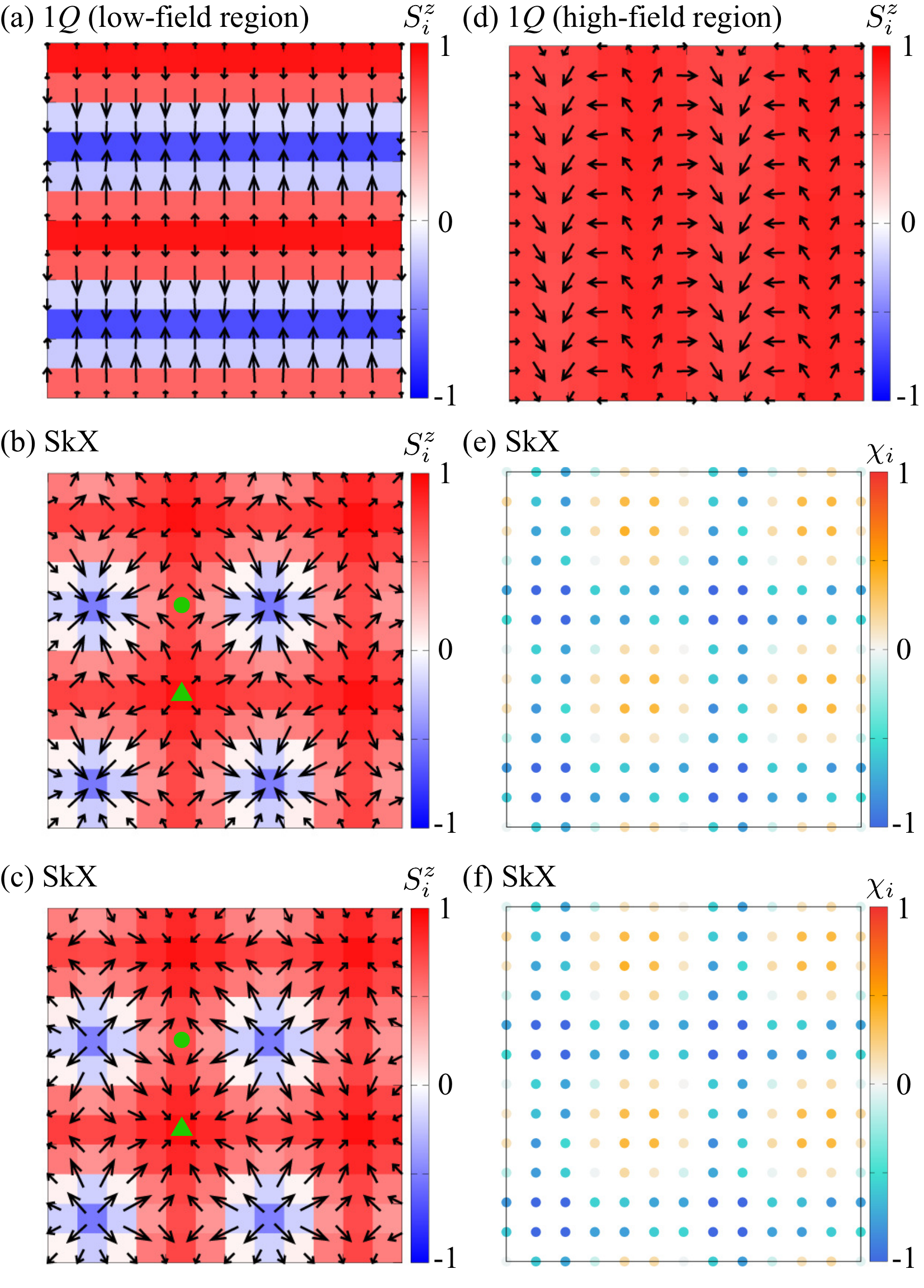} 
\caption{
\label{Fig: SL_spin}
Real-space spin configurations of (a) the 1$Q$ state in the low-field region at $H=0.4$, (b,c) the SkX at $H=0.8$, and (d) the 1$Q$ state in the high-field region at $H=1.2$ for $J_{\parallel}=0$. 
The arrows represent the $xy$ components of the spin moment and the color shows the $z$ component. 
The data in (a), (b), and (d) are obtained for $D=0.2$, while those in (c) are obtained for $D=-0.2$. 
In (b) and (c), the triangle and circle represent two types of vortex cores (vortex-1 core and vortex-2 core); see the main text in detail. 
(e, f) Real-space scalar chirality configurations corresponding to the spin configurations of (b) and (c) are shown. 
}
\end{center}
\end{figure*}

\begin{figure*}[htb!]
\begin{center}
\includegraphics[width=1.0 \hsize]{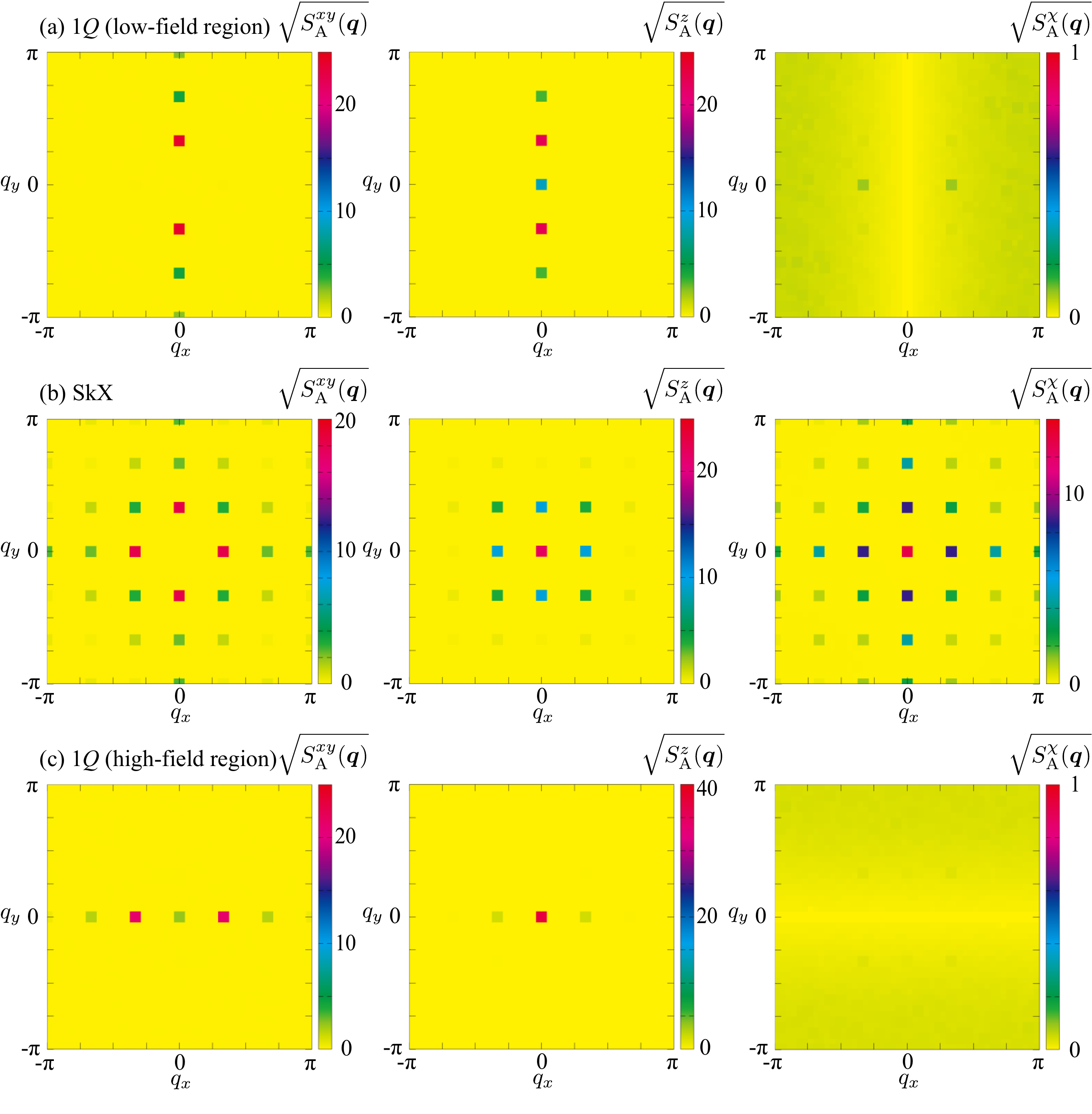} 
\caption{
\label{Fig: Sq_SL}
(Left and middle) The square root of the $xy$ and $z$ components of the spin structure factor in the first Brillouin zone for layer A in (a) the 1$Q$ state in the low-field region at $H=0.4$, (b) the SkX at $H=0.8$, and (c) the 1$Q$ state in the high-field region at $H=1.2$ for $J_{\parallel}=0$.
(Right) The square root of the chirality structure factor for layer A. 
}
\end{center}
\end{figure*}

We consider the instability of the SkX and the spiral state in the single-layer system by setting $J_{\parallel}=0$; hereafter, we present the result for layer A in this section. 
Then, the system is regarded as the noncentrosymmetric polar system with a uniform DM interaction. 
Figures~\ref{Fig: Mag_chirality_SL}(a) and \ref{Fig: Mag_chirality_SL}(b) show the $H$ dependences of the magnetization $M_{\rm A}^z$ and the scalar chirality $\chi_{\rm A}^{\rm sc}$ at $D=0.2$, respectively. 
Although both quantities become zero at $H=0$, $M_{\rm A}^z$ gradually increases while increasing $H$.
Meanwhile, $\chi_{\rm A}^{\rm sc}$ remains zero for small $H$. 
The real-space spin configuration for small $H=0.4$ is shown in figure~\ref{Fig: SL_spin}(a), which is characterized by the out-of-plane cycloidal spiral modulation along the $\bm{Q}_2$ direction; the spiral plane lies on the $xz$ plane. 
We refer to this spin state as a 1$Q$ state. 
The 1$Q$ state exhibits the single-$Q$ peak structure at $\bm{Q}_2$ in the spin structure factor, as shown in the left and middle panels of figure~\ref{Fig: Sq_SL}(a).
The additional structures at higher harmonic wave vectors like $2\bm{Q}_2$ and $3\bm{Q}_2$ appear in the presence of the magnetic field, which modulates the spiral plane from the circular shape to the elliptical shape. 
It is noted that the energetically degenerate spiral state with the peak structure at $\bm{Q}_1$ instead of $\bm{Q}_2$ is obtained in the simulations depending on the initial spin configuration due to the fourfold rotational symmetry. 
Reflecting the coplanar spin structure, the 1$Q$ state does not show a peak structure in the chirality structure factor, as shown in the right panel of figure~\ref{Fig: Sq_SL}(a). 

While a further increase of $H$, the 1$Q$ state turns into another magnetically ordered state at $H \simeq 1.05$ while showing jumps of $M_{\rm A}^z$ and $\chi_{\rm A}^{\rm sc}$ in figure~\ref{Fig: Mag_chirality_SL}; both quantities become nonzero for $1.05 \lesssim H \lesssim 1.25$, which indicates the emergence of the SkX. 
In this intermediate field region, one finds the square lattice formation of the skyrmion core at $S_i^z=-1$ in the real-space picture in figure~\ref{Fig: SL_spin}(b). 
The skyrmion core is located at the center of the square plaquette due to the nature of the discrete lattice system~\cite{Hayami_PhysRevResearch.3.043158}, which is surrounded by spins parallel to the radial directions so as to have the positive winding number of $+1$. 
This indicates that this SkX is identified as the N\'eel SkX with $n^{(\rm A)}_{\rm sk}=-1$. 
It is noted that this SkX includes the other two types of vortex cores in addition to the skyrmion cores: One is the vortex with the winding number of $+1$ surrounded by the four skyrmion cores and the other is that with the winding number of $-1$ surrounded by the two skyrmion cores, as denoted by the triangle and circle in figure~\ref{Fig: SL_spin}(b), respectively. 
We call the former a vortex-1 core and the latter a vortex-2 core. 
By considering the $z$-spin polarization in each core, the local scalar chirality becomes positive around the vortex-1 core, while that becomes negative around the vortex-2 core, as shown in figure~\ref{Fig: SL_spin}(e).

From the momentum-space viewpoint, the SkX spin texture is characterized by a double-$Q$ superposition of two cycloidal spiral waves at $\bm{Q}_1$ and $\bm{Q}_2$, as shown in the left and middle panels of figure~\ref{Fig: Sq_SL}(b). 
In addition, this state also accompanies the chirality density waves at the finite $\bm{Q}$ components in addition to the uniform ($\bm{q}=\bm{0}$) component; the dominant intensities are found at $\bm{Q}_1$ and $\bm{Q}_2$ in the chirality structure factor as shown in the right panel of figure~\ref{Fig: Sq_SL}(c). 

It is noted that the spin configuration of the SkX depends on the sign of the DM interaction, although its stability region remains the same. 
For example, we show the spin configuration in the SkX phase when setting $D=-0.2$ and $H=1.2$ in figure~\ref{Fig: SL_spin}(c). 
The sign of the $xy$-spin components is opposite compared to that of the spin configuration in figure~\ref{Fig: SL_spin}(b); the direction of the in-plane spins around the skyrmion core is inward in figure~\ref{Fig: SL_spin}(b), while that is outward in figure~\ref{Fig: SL_spin}(c). 
As a result, the helicities of the skyrmion core, vortex-1 core, and vortex-2 core become opposite. 
Meanwhile, the topological property is unchanged when reversing the sign of the DM interaction, since the sign of the scalar chirality is unchanged, as shown by the real-space chirality configuration in figures~\ref{Fig: SL_spin}(e) and \ref{Fig: SL_spin}(f). 

When the magnetic field is increased from the SkX phase, the 1$Q$ state with zero $\chi^{\rm sc}_{{\rm A}}$ appears again in the high-field region, as shown in figure~\ref{Fig: Mag_chirality_SL}. 
Similar to the 1$Q$ state in the low-field region, the spiral state with the peak structure at $\bm{Q}_1$ has the same energy as that at $\bm{Q}_2$; in the simulations, their appearance depends on the initial spin configuration. 
In contrast to the 1$Q$ state in the low-field region, the spiral plane is tilted from the out-of-plane cycloidal plane to the in-plane one so as to gain the Zeeman energy. 
Accordingly, the modulation regarding the $z$-spin component becomes small, as found in the real-space spin configuration in figure~\ref{Fig: SL_spin}(d) and the spin structure factor in the middle panel of figure~\ref{Fig: Sq_SL}(c). 
Besides, the tilted spiral spin configuration induces the in-plane magnetization, as shown in the left panel of figure~\ref{Fig: Sq_SL}(c). 
This state does not exhibit the chirality density wave similar to the low-field case, as shown in the right panel of figure~\ref{Fig: Sq_SL}(c). 
This 1$Q$ state continuously turns into the fully-polarized state at $H=2$ shown in figure~\ref{Fig: Mag_chirality_SL}(a). 
The resultant phase sequence against $H$ is consistent with that in the model with the FM exchange interaction and the DM interaction between the nearest-neighbor spins~\cite{Yi_PhysRevB.80.054416, Mochizuki_PhysRevLett.108.017601, Rowland_PhysRevB.93.020404}. 

\subsection{Bilayer case}
\label{sec: bilayer}

\begin{figure}[htb!]
\begin{center}
\includegraphics[width=1.0 \hsize]{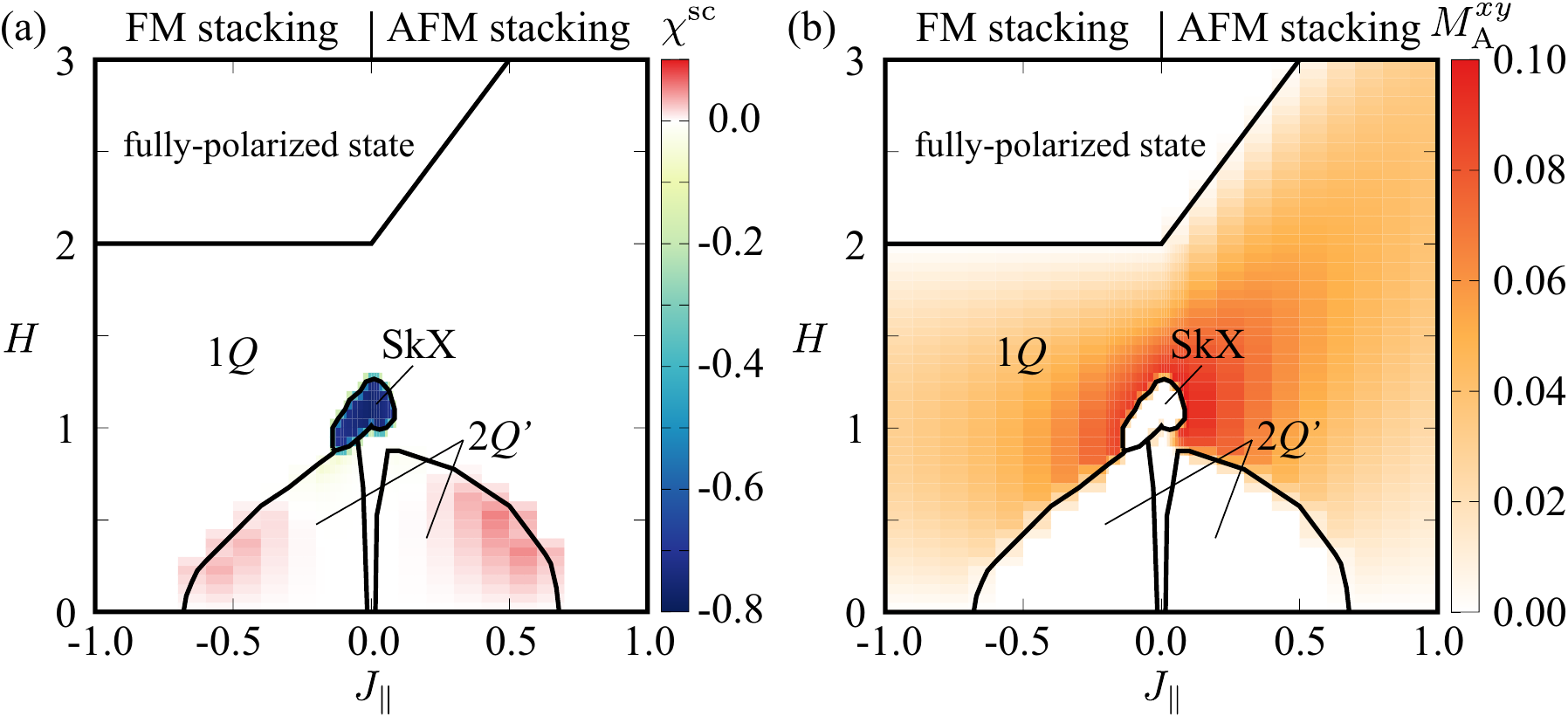} 
\caption{
\label{Fig: PD}
(a,b) The phase diagram in the plane of the interlayer exchange interaction $J_{\parallel}$ and the magnetic field $H$. 
The color plot represents (a) the spin scalar chirality, $\chi^{\rm sc}$, and (b) the in-plane magnetization for layer A, $M_{\rm A}^{xy}$. 
The regions for $J_{\parallel}>0$ and $J_{\parallel}<0$ represent the cases of the antiferromagnetic (AFM) and ferromagnetic (FM) interlayer interactions, respectively. 
}
\end{center}
\end{figure}

\begin{figure*}[htb!]
\begin{center}
\includegraphics[width=1.0 \hsize]{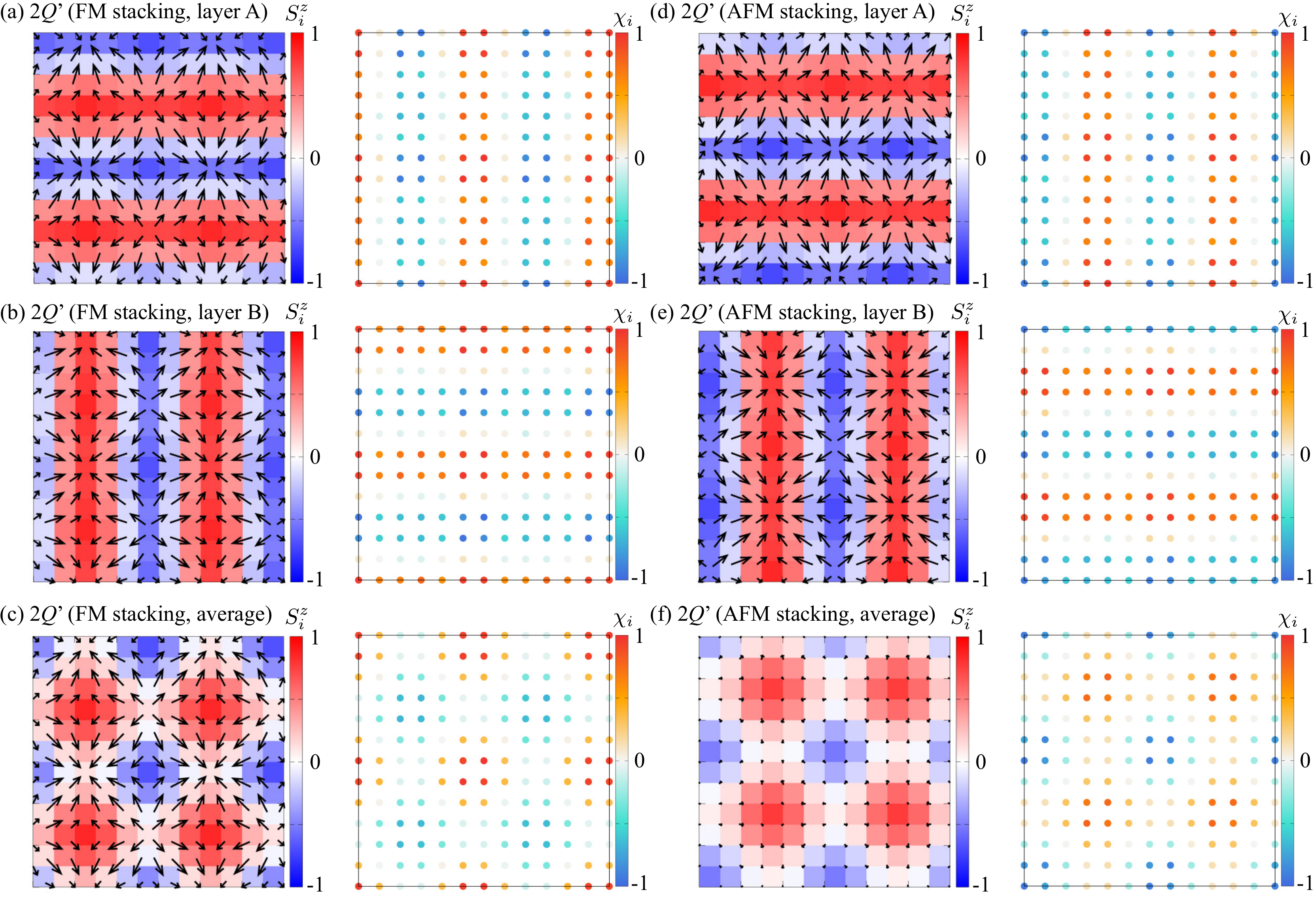} 
\caption{
\label{Fig: 2Q}
(Left) Real-space spin configurations of the 2$Q'$ state on (a), (d) layer A and (b), (e) layer B in (a), (b) the FM stacked case at $J_{\parallel}=-0.5$ and $H=0.4$ and (d), (e) the AFM stacked case at $J_{\parallel}=0.5$ and $H=0.5$. 
(c), (f) The averaged spin configurations over layers A and B in (c) the FM stacked case and (f) the AFM stacked case. 
The arrows represent the $xy$ components of the spin moment and the color shows the $z$ component. 
(Right) Real-space scalar chirality configurations corresponding to the spin configuration. 
}
\end{center}
\end{figure*}

\begin{figure*}[htb!]
\begin{center}
\includegraphics[width=1.0 \hsize]{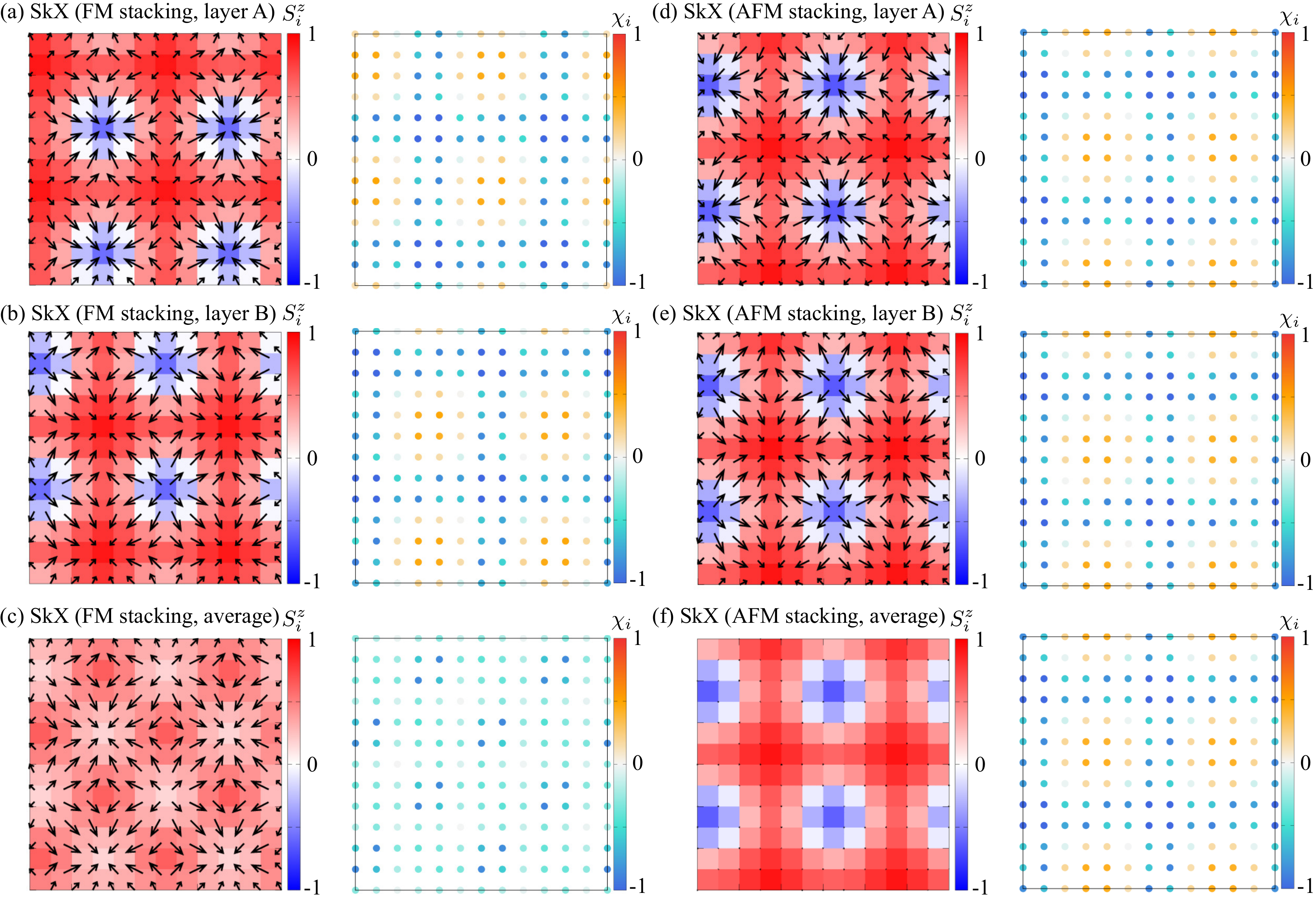} 
\caption{
\label{Fig: SkX}
(Left) Real-space spin configurations of the SkXs on (a), (d) layer A and (b), (e) layer B in (a), (b) the FM stacked case at $J_{\parallel}=-0.1$ and $H=1$ and (d), (e) the AFM stacked case at $J_{\parallel}=0.05$ and $H=1$. 
(c), (f) The averaged spin configurations over layers A and B in (c) the FM stacked case and (f) the AFM stacked case. 
The arrows represent the $xy$ components of the spin moment and the color shows the $z$ component. 
(Right) Real-space scalar chirality configurations corresponding to the spin configuration. 
}
\end{center}
\end{figure*}

\begin{figure*}[htb!]
\begin{center}
\includegraphics[width=1.0 \hsize]{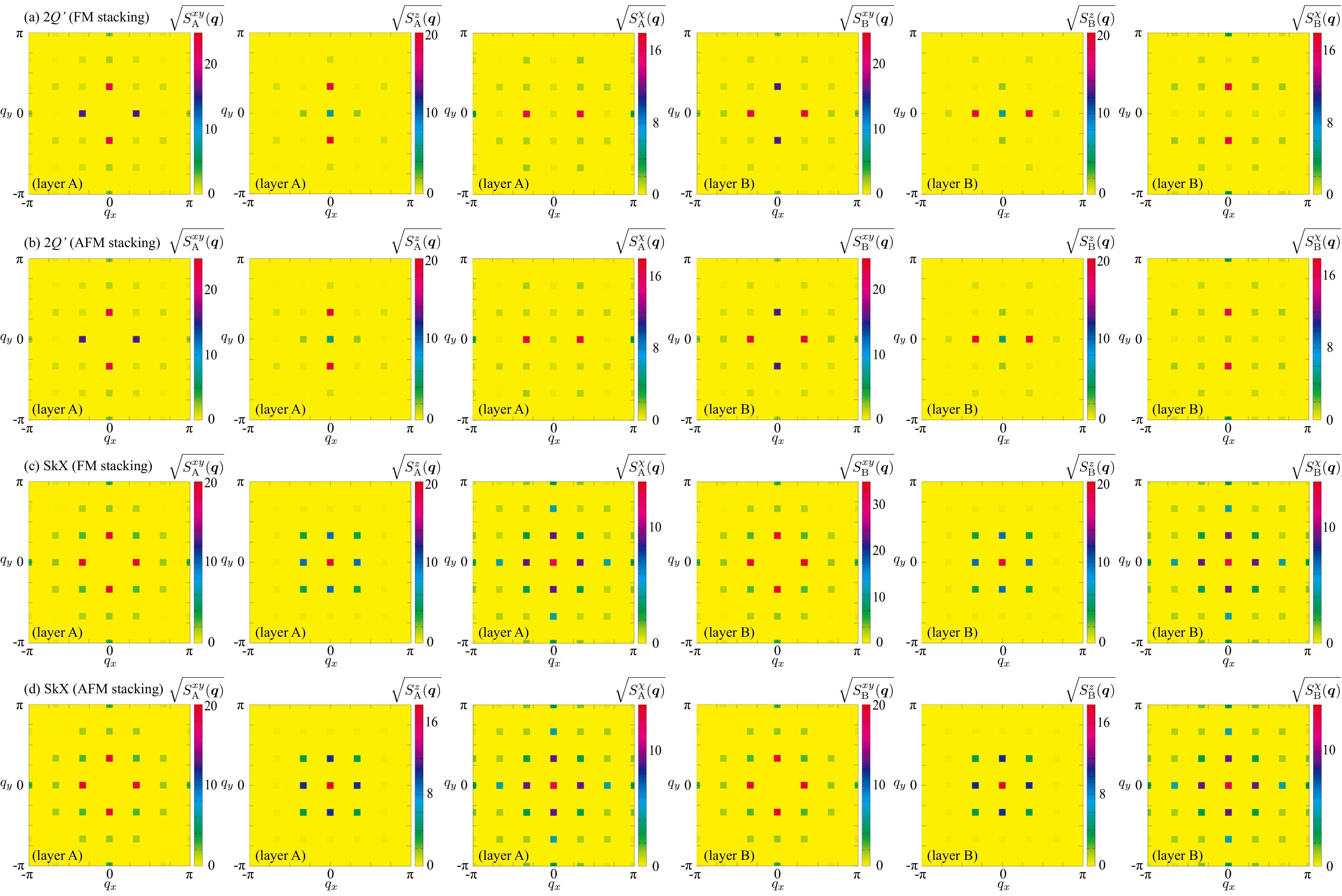} 
\caption{
\label{Fig: Sq}
(Left and second left) The square root of the $xy$ and $z$ components of the spin structure factor in the first Brillouin zone for layer A in (a) the 2$Q'$ state at $J_{\parallel}=-0.5$ and $H=0.4$, (b) the 2$Q'$ state at $J_{\parallel}=0.5$ and $H=0.5$, (c) the SkX at $J_{\parallel}=-0.1$ and $H=1$, and (d) the SkX at $J_{\parallel}=0.05$ and $H=1$.
(Middle left) The square root of the chirality structure factor for layer A. 
The right three panels represent the data for layer B corresponding to the left three ones. 
}
\end{center}
\end{figure*}

Next, we consider the effect of the interlayer exchange interaction $J_{\parallel}$ for the bilayer square-lattice model with the staggered DM interaction. 
Figure~\ref{Fig: PD} shows the low-temperature phase diagram at $T=0.001$, which is obtained by simulated annealing while changing the interlayer exchange coupling $J_{\parallel}$ and the magnetic field $H$ on the bilayer square lattice. 
The region for $J_{\parallel}<0$ ($J_{\parallel}>0$) corresponds to the FM (AFM) interlayer interaction. 
The contour plots of figures~\ref{Fig: PD}(a) and \ref{Fig: PD}(b) represent the total scalar chirality, $\chi^{\rm sc}$, and the inplane magnetization for layer A, $M_{\rm A}^{xy}$, respectively. 
When $J_{\parallel}=0$, the 1$Q$ state, the SkX, the 1$Q$ state, and the fully-polarized state appears in order upon increasing $H$, as shown in Sec.~\ref{sec: single layer}. 

In the following, we discuss the details of the spin and chirality configurations in the 2$Q'$ state stabilized in the low-field region and in the SkX in the intermediate-field region, both of which exhibit a uniform scalar chirality, as shown in Fig.~\ref{Fig: PD}(a).  
Meanwhile, we omit the result of the other 1$Q$ state in figure~\ref{Fig: PD}(a), since it shows a similar spin configuration to that in the single-layer model in figure~\ref{Fig: SL_spin}(d); the spin structure factor shows a dominant intensity at $\bm{Q}_1$ or $\bm{Q}_2$ in addition to the in-plane magnetization at $\bm{q}=\bm{0}$. 
We show a contour plot of $M_{\rm A}^{xy}$ in figure~\ref{Fig: PD}(b), where the appearance of $M_{\rm A}^{xy}$ corresponds to the region of the 1$Q$ state except for small $J_{\parallel}$ in the low-field region. 
It is noted that the in-plane magnetization is induced in a staggered way for different layers so as to vanish the net magnetization.  
The 1$Q$ state changes into the fully-polarized state at $H=2$ for $J_{\parallel}<0$ and at $H=2(1+J_{\parallel})$ for $J_{\parallel}>0$, as we ignore the effect of the interaction at $\bm{q}=\bm{0}$. 

The snapshots of the real-space spin and chirality configurations obtained by simulated annealing in two phases (2$Q'$ state and SkX) are shown in figures~\ref{Fig: 2Q} and \ref{Fig: SkX}. 
In addition, we also show the spin and chirality structure factors in momentum space in figure~\ref{Fig: Sq}. 
In both real and momentum spaces, we show the spin- and chirality-related quantities in each layer. 
Furthermore, we show the averaged spin and chirality configurations over layers in order to clearly show the similarity and difference between layers A and B. 

First, we discuss the 2$Q'$ state, which is realized in the low-field region by taking into account $J_{\parallel} \neq 0$, as shown in figure~\ref{Fig: PD}(a). 
This state is stabilized for both FM and AFM interlayer interactions, whose stability region is almost symmetric in terms of the FM and AFM interlayer interactions. 

The real-space spin configurations in the 2$Q'$ state are presented in the case of the FM interlayer interaction in the left panel of figures~\ref{Fig: 2Q}(a) and \ref{Fig: 2Q}(b) and the case of the AFM interlayer interaction in the left panel of figures~\ref{Fig: 2Q}(d) and \ref{Fig: 2Q}(e), which are modulated from the single-$Q$ cycloidal spiral wave in the single-layer 1$Q$ state in figure~\ref{Fig: SL_spin}(a). 
A way of modulation is found in the spin structure factor in figures~\ref{Fig: Sq}(a) and \ref{Fig: Sq}(b); in each layer, there are dominant single-$Q$ peaks at either $\bm{Q}_1$ or $\bm{Q}_2$ in both $xy$ and $z$ spin components and the subdominant peaks at the remaining other $\bm{Q}_\nu$. 
In other words, the intensities of the spin structure factor at $\bm{Q}_1$ and $\bm{Q}_2$ are different from each other, although the fourfold rotational symmetry seems to be recovered by summing the intensities in the spin structure factor over the layers, i.e., $S^{xy}_{\rm A}(\bm{Q}_1)+S^{xy}_{\rm B}(\bm{Q}_1)=S^{xy}_{\rm A}(\bm{Q}_2)+S^{xy}_{\rm B}(\bm{Q}_2)$ and $S^{z}_{\rm A}(\bm{Q}_1)+S^{z}_{\rm B}(\bm{Q}_1)=S^{z}_{\rm A}(\bm{Q}_2)+S^{z}_{\rm B}(\bm{Q}_2)$. 
This double-$Q$ superposition leads to a noncoplanar spin configuration with nonzero local scalar chirality, as shown in the right panel of figures~\ref{Fig: 2Q}(a), \ref{Fig: 2Q}(b), \ref{Fig: 2Q}(d), and \ref{Fig: 2Q}(e). 
Indeed, the chirality structure factor $S^{\chi}_{\eta}(\bm{q})$ shows the dominant peak at the second-largest $\bm{Q}_\eta$ component in the spin structure factor. 
For example, one finds a chirality density wave along the $x$ direction in real space ($\bm{Q}_1$ component) perpendicular to the dominant spiral modulation along the $y$ direction ($\bm{Q}_2$ component) in figure~\ref{Fig: 2Q}(a). 
A similar double-$Q$ state with the chirality density wave has been discussed in itinerant magnets without the spin-orbit coupling on a single-layer square lattice~\cite{Solenov_PhysRevLett.108.096403, Ozawa_doi:10.7566/JPSJ.85.103703,yambe2020double}, where the multiple-spin interactions arising from the itinerant nature of electrons play an important role. 
Meanwhile, the present 2$Q'$ state is stabilized by incorporating the effect of both the staggered DM and interlayer interactions. 
In fact, the 2$Q'$ state turns into the 1$Q$ state when decreasing $D$ (not shown) or $J_{\parallel}$. 
Thus, the 2$Q'$ state is a consequence of the bilayer lattice system with the staggered DM interaction. 

Notably, the $\bm{q}$-peak structure shows the layer dependence in the 2$Q'$ state irrespective of the FM and AFM interactions, as shown in figures~\ref{Fig: Sq}(a) and \ref{Fig: Sq}(b); the dominant peak of $S^{xy}_{\rm A}(\bm{q})$ and $S^{z}_{\rm A}(\bm{q})$ lies at $\bm{Q}_2$, while that of $S^{xy}_{\rm B}(\bm{q})$ and $S^{z}_{\rm B}(\bm{q})$ lies at $\bm{Q}_1$. 
Similarly, the chirality structure factor also shows the layer-dependent structure; the dominant peak of $S^{\chi}_{\rm A}(\bm{q})$ is found at $\bm{Q}_1$, while that of $S^{\chi}_{\rm B}(\bm{q})$ is found at $\bm{Q}_2$. 
One finds such a difference in the real-space spin and chirality configurations in figures~\ref{Fig: 2Q}(a), \ref{Fig: 2Q}(b), \ref{Fig: 2Q}(d), and \ref{Fig: 2Q}(e). 
This layer-dependent $\bm{q}$ peak structure is attributed to the staggered DM interaction that fixes the helicity of the spiral in an opposite way for layers A and B, which brings about magnetic frustration. 
A similar situation also happens for the bilayer triangular-lattice model~\cite{Hayami_PhysRevB.105.014408}. 
The difference between the FM and AFM interlayer interactions appears in the relative phase difference for layers A and B; the spins for different layers are aligned so as to gain the energy by $J_{\parallel}$ in terms of the $xy$-spin component, as clearly found in the averaged spin configuration over layers A and B in figures~\ref{Fig: 2Q}(c) and \ref{Fig: 2Q}(f). 
This is because the intensity of $S^{xy}_{\eta}(\bm{Q}_\nu)$ is larger than that of $S^{z}_{\eta}(\bm{Q}_\nu)$, as shown in figures~\ref{Fig: Sq}(a) and \ref{Fig: Sq}(b). 
Such a situation might change when introducing the easy-axis single-ion anisotropy like $-A \sum_i (S_i^z)^2$ leading to $S^{xy}_{\eta}(\bm{Q}_\nu) < S^{z}_{\eta}(\bm{Q}_\nu)$; it is expected that the spins for different layers are aligned so as to gain the energy by $J_{\parallel}$ in terms of the $z$-spin component, although a further investigation is required. 

Owing to the noncoplanar double-$Q$ spin texture, this state exhibits the uniform component of the scalar chirality $\chi^{\rm sc}$ in addition to the finite-$Q$ component, as shown in figure~\ref{Fig: PD}; see also the $\bm{q}=\bm{0}$ component of the chirality structure factor in figures~\ref{Fig: Sq}(a) and \ref{Fig: Sq}(b). 
The sign of the induced scalar chirality depends on the magnitude of $J_{\parallel}$; it tends to become negative (positive) for small (large) $|J_{\parallel}|$ in the $2Q'$ phase, although the negative value is much smaller compared to the SkX in figure~\ref{Fig: PD}(a). 
For example, $\chi^{\rm sc} \simeq -0.0497$ at $H=0.85$, $\chi^{\rm sc} \simeq -0.0015$ at $H=0.5$, and $\chi^{\rm sc} \simeq -0.00018$ at $H=0.3$ in the case of $J_{\parallel}=-0.1$. 
It is noted that the small value of $\chi^{\rm sc}$ is not due to the finite-size effect.
The nonzero chirality is owing to an imbalance between the regions with the positive and negative chiralities in real space, as shown in the right panel of figure~\ref{Fig: 2Q}. 
Thus, the anomalous Hall effect triggered by the noncoplanar spin texture is expected in the 2$Q'$ state. 
On the other hand, it is noted that the skyrmion number is not quantized in the 2$Q'$ state, which indicates that this state is topologically trivial in contrast to the SkX. 
Since the magnitude of $\chi^{\rm sc}$ becomes larger while increasing $|J_{\parallel}|$, the layered structure with the different ordering vectors is an essence to induce $\chi^{\rm sc}$. 
A similar multiple-$Q$ state with a nonzero scalar chirality in the low-field region has also been found in the bilayer triangular-lattice model~\cite{Hayami_PhysRevB.105.014408}. 
Thus, the emergence of such a state with a nonzero scalar chirality is one of the characteristic points in the bilayer system with the staggered DM interaction irrespective of the detailed lattice structures.

Next, we discuss the square SkX appearing in the narrow region for small $|J_{\parallel}|$ in the intermediate field in figure~\ref{Fig: PD}(a). 
Although the square SkX is stabilized for both FM and AFM interlayer interactions, the stability range is different from each other. 
For the FM case, the square SkX is stabilized up to $J_{\parallel} \simeq -0.14$, while it is stabilized up to $J_{\parallel} \simeq 0.085$ for the AFM case. 
Since the square SkX phase for nonzero $J_{\parallel}$ smoothly connects that at  $J_{\parallel}=0$, its origin is attributed to the presence of the DM interaction. 
This result clearly indicates that the staggered DM interaction can become the microscopic origin of the square SkX in centrosymmetric magnets in addition to the triangular SkX~\cite{Hayami_PhysRevB.105.014408,lin2021skyrmion}.

\begin{figure}[htb!]
\begin{center}
\includegraphics[width=1.0 \hsize]{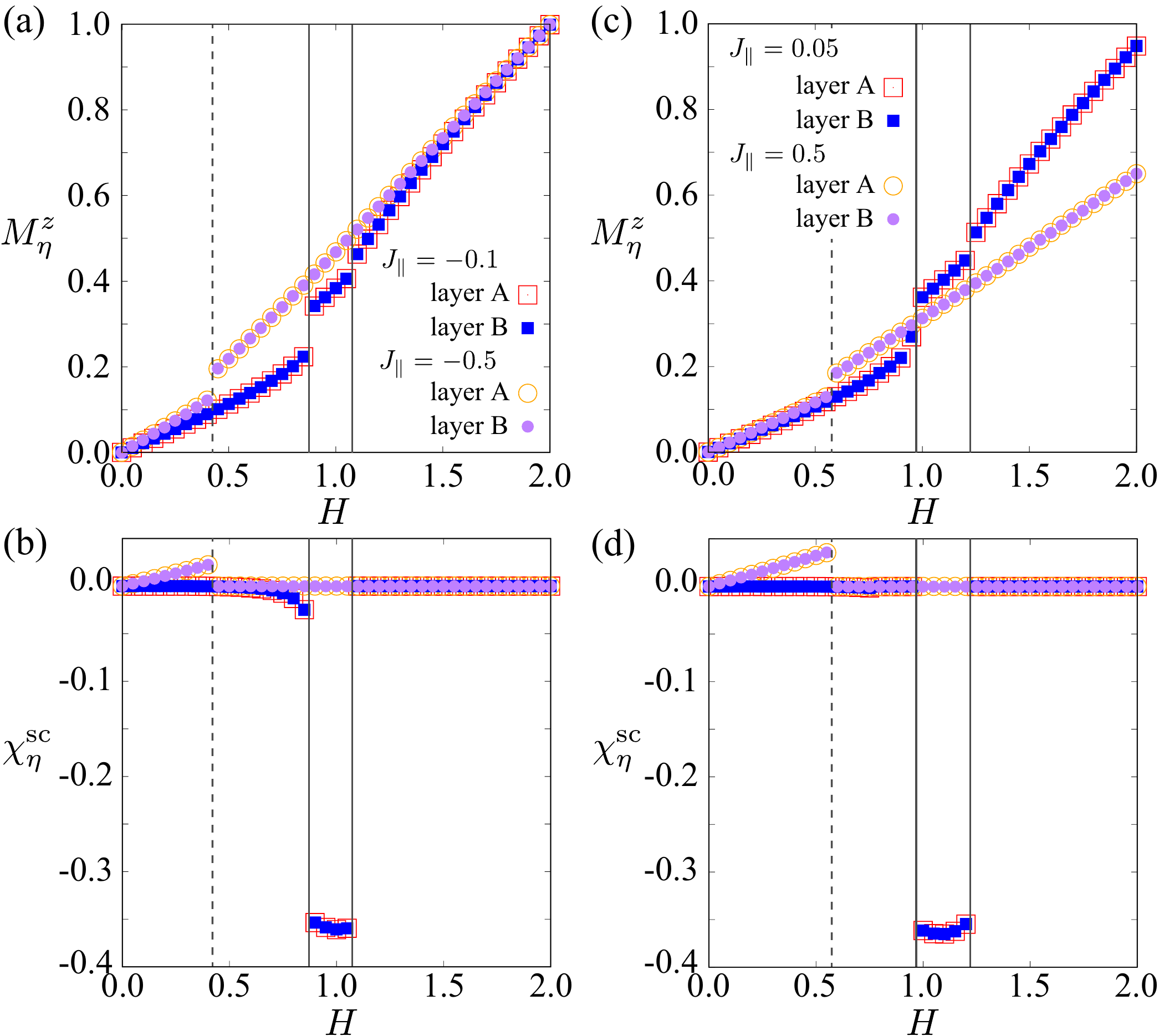} 
\caption{
\label{Fig: Mag_chirality}
$H$ dependences of (a), (c) the magnetization $M^z_\eta$ and (b), (d) the scalar chirality $\chi^{\rm sc}_\eta$ for layers $\eta=$A and B at (a), (b) $J_{\parallel}=-0.1$ and $-0.5$ and (c), (d) $J_{\parallel}=0.05$ and $0.5$. 
The solid (dashed) vertical lines represent the phase boundaries for small (large) $|J_{\parallel}|$. 
}
\end{center}
\end{figure}

The SkX phase consists of the SkX layers with different helicities for both FM and AFM interlayer interactions, as shown in figures~\ref{Fig: SkX}(a), \ref{Fig: SkX}(b), \ref{Fig: SkX}(d), and \ref{Fig: SkX}(e). 
The SkX spin and chirality textures in real space for each layer well correspond to that obtained in the single-layer model, as shown in figures~\ref{Fig: SL_spin}(b) and \ref{Fig: SL_spin}(c). 
The behaviors of the spin and chirality structure factors in the bilayer case in figures~\ref{Fig: Sq}(c) and \ref{Fig: Sq}(d) are also similar to those in the single-layer case in figure~\ref{Fig: Sq_SL}(b). 
The skyrmion number in each layer is quantized as $-1$.

Reflecting the different helicities of the SkX in each layer, the relative skyrmion core positions are different for FM and AFM interlayer interactions, which is clearly found in local spin and chirality configurations in real space shown in figure~\ref{Fig: SkX}. 
For the FM interlayer interaction, the skyrmion cores lie at the different positions on layers A and B as shown in figures~\ref{Fig: SkX}(a) and \ref{Fig: SkX}(b), while those lie at the same position for the AFM interlayer interaction as shown in figures~\ref{Fig: SkX}(d) and \ref{Fig: SkX}(e). 
Specifically, the skyrmion core on layer B lies at the same position as the vortex-1 core on layer A in the FM interlayer interaction. 
This difference is understood from the energetic point of view; the SkXs are stacked so as to align the in-plane spins on two layers in a (anti)parallel way in the FM (AFM) interlayer interaction to gain the exchange energy in terms of the $xy$ spin component rather than the $z$ spin component. 
Indeed, the intensities of $xy$ spin in the spin structure factor at $\bm{Q}_\nu$ component are larger than those of $z$ spin. 
Such a difference regarding the skyrmion core position between the FM and AFM interlayer interactions is found in the averaged spin textures over the layers in figures~\ref{Fig: SkX}(c) and \ref{Fig: SkX}(f).

Finally, let us discuss the phase transition between the obtained phases in terms of the magnetization $M^{z}_\eta$ and the scalar chirality $\chi_{\eta}^{\rm sc}$ in figure~\ref{Fig: Mag_chirality}. 
Figures~\ref{Fig: Mag_chirality}(a) and \ref{Fig: Mag_chirality}(b) represent the $H$ dependences of $M^{z}_\eta$ and $\chi_{\eta}^{\rm sc}$, respectively, for $J_{\parallel}=-0.1$ and $-0.5$. 
For $J_{\parallel}=-0.1$, the phase transitions between the 2$Q'$ state and the SkX and between the SkX and $1Q$ state denoted by the solid lines are characterized by the first-order phase transition with jumps of $M^{z}_\eta$ [figure~\ref{Fig: Mag_chirality}(a)] and $\chi_{\eta}^{\rm sc}$ [figure~\ref{Fig: Mag_chirality}(b)]. 
Similarly, the phase transition between the 2$Q'$ state and the 1$Q$ state denoted by the dashed lines is of first order with a jump of $M^z_\eta$ [figure~\ref{Fig: Mag_chirality}(a)]. 
It is noted that the scalar chirality in the 2$Q'$ state takes negative (positive) values for small (large) $|J_{\parallel}|$, as discussed above. 
A similar tendency is found in the case of the AFM interlayer interaction, as shown in figures~\ref{Fig: Mag_chirality}(c) and \ref{Fig: Mag_chirality}(d).

\section{Summary}
\label{sec: summary}

To summarize, we have investigated the possibility of the square SkX in the centrosymmetric tetragonal system with the bilayer structure. 
By carrying out the simulated annealing for the spin model, we found that the staggered DM interaction that originates from the bilayer structure is a microscopic key ingredient to stabilizing the square SkX at low temperatures. 
The obtained SkX remains stable for both FM and AFM interlayer exchange interactions, although their stability region becomes narrower compared to the bilayer triangular-lattice model. 
We also show that the single-$Q$ spiral state in the low-field region in the single-layer model is replaced with the double-$Q$ state with a net scalar chirality by taking into account the interlayer exchange interaction. 
The present results provide another way to realize the square SkX in centrosymmetric magnets with the sublattice degree of freedom. 
As the recent studies have indicated the possibility of further exotic SkXs in the multi-sublattice systems, such as the AFM SkX~\cite{Rosales_PhysRevB.92.214439,zhang2016antiferromagnetic, Gobel_PhysRevB.96.060406, Kravchuk_PhysRevB.99.184429,gao2020fractional, Tome_PhysRevB.103.L020403}, and their related dynamics~\cite{zhang2016magnetic, Zhang_PhysRevB.94.064406,koshibae2017theory,hrabec2017current, Shen_PhysRevB.98.134448,ang2019bilayer, Xia_PhysRevApplied.11.044046}, it is intriguing to explore a further possibility to realize the AFM SkX based on the present layered model with the layer-dependent DM interaction, which will be left for future study.

\ack
This research was supported by JSPS KAKENHI Grants Numbers JP21H01037, JP22H04468, JP22H00101, JP22H01183 and by JST PRESTO (JPMJPR20L8). 
Parts of the numerical calculations were performed in the supercomputing systems in ISSP, the University of Tokyo.

\vspace{8mm}

\bibliographystyle{iopart-num}
\bibliography{ref}

\end{document}